\documentclass[letterpaper,showkeys,showpacs,preprintnumbers,eqsecnum,superscriptaddress,aps,nofootinbib]{revtex4}
\usepackage{amssymb}
\usepackage[centertags]{amsmath}
\usepackage{txfonts}
\usepackage{epsfig}
\usepackage{bm}
\usepackage{color} 
\usepackage{graphicx,graphics}\usepackage{multirow}\usepackage{float}\usepackage{ulem}
\usepackage[unicode=true,pdfusetitle]{hyperref}
\usepackage{setspace}\usepackage{slashed}
\usepackage{mathrsfs}

\newcommand{\jp}[3]{\frac{#1}{#2}^{#3}}
\newcommand{\im}{\mathrm{i}}

\begin{document}

\title{Tensor amplitudes for partial wave analysis of \texorpdfstring{$\psi \to\Delta\bar{\Delta}$}{} within helicity frame}

\author{Xiang Dong}
\affiliation{Wuhan University, Wuhan 430072, People's Republic of China}

\author{Kexin Su}
\affiliation{Wuhan University, Wuhan 430072, People's Republic of China}

\author{Hao Cai}
\email{hcai@whu.edu.cn}
\affiliation{Wuhan University, Wuhan 430072, People's Republic of China}

\author{Kai Zhu}
\email{zhuk@ihep.ac.cn}
\affiliation{Institute of High Energy Physics, Beijing 100049, China}

\author{Yonggui Gao}
\affiliation{Institute of High Energy Physics, Beijing 100049, China}

\begin{abstract}
We have derived the tensor amplitudes for partial wave analysis of $\psi\to\Delta\bar{\Delta}$, $\Delta \to p \pi$ within the helicity frame, as well as the amplitudes for the other decay sequences with same final states. These formulae are practical for the experiments measuring $\psi$ decaying into $p\bar{p}\pi^+ \pi^-$ final states, such as BESIII with its recently collected huge $J/\psi$ and
$\psi(2S)$ data samples.

\end{abstract}

\maketitle

\section{Introduction}

Recently, many measurements of $J/\psi$ or $\psi(2S)$ decays into baryon-pairs are
reported by BESIII collaboration, such as $\psi(2S) \to \Omega^-
\bar{\Omega}^+$~\cite{BESIII:2020lkm}, $\psi(2S) \to p \bar{p} (n\bar{n})$~\cite{1803.02039},
$J/\psi \left(\psi(2S)\right) \to \Sigma^+ \bar{\Sigma}^-$~\cite{2004.07701},
$J/\psi \left(\psi(2S)\right) \to \Xi^0 \bar{\Xi}^0
$~\cite{1612.08664}, $J/\psi \left(\psi(2S)\right) \to \Xi^- \bar{\Xi}^+
$~\cite{1602.06754}, $J/\psi \to \Xi(1530)^- \bar{\Xi}^+$~\cite{1911.06669}, $\psi(2S) \to
\Xi(1530)^- \Xi(1530)^+ (\Xi(1530)^- \Xi^+)$~\cite{1907.13041} etc. These measurements are
based on the huge $J/\psi$ or $\psi(2S)$ samples collected at the BESIII detector, who
is operating at the electron position collider BEPCII. Taking advantage of the improved
statistics, more precise measurements have been achieved and some insights on the physics
have been provided, such as the polarization parameters of baryons, relative phase between
electric-magnetic and strong amplitudes, decay mechanism of $\psi$ decaying into baryon
pairs, and search for excited baryon states, etc. However, there is no measurement of
$\psi \to \Delta \bar{\Delta}$ among these new exciting experimental results. A single
$\psi$ will represent both $J/\psi$ and $\psi(2S)$ states later in this paper if
not specified. The latest measurement of $\psi \to \Delta \bar{\Delta}$ is done by BESII
about 20 years ago~\cite{Bai:2000ye}. The lack of $\Delta \bar{\Delta}$ study does
not mean this channel is not essential. In contrast, a careful analysis of $\psi \to \Delta
\bar{\Delta}$ would provide a lot of important information of the decay mechanism of the
vector charmonia to the pair of $\Delta$, the first discovered resonance in particle physics, as
well as the line-shape of the invariant mass of the final states of this resonance. The main difficulty of
this measurement is that $\Delta$ has a much broader width than other
baryon states. From PDG~\cite{Zyla:2020zbs}, the Breit-Wigner width of the
$\Delta(1232)^{++}$ is about 117 MeV, the widths of other excited $\Delta$ states are at
the same level. It makes the interference effect significant in the measurement,
but that has been ignored in previous
measurements~\cite{Bai:2000ye,Eaton:1983kb}. When we try to measure $\psi \to \Delta \bar{\Delta}$
via final states protons and pions, that are the products of the strong decays of $\Delta$, the interference between
$\Delta^{++} \Delta^{--}$ and $\Delta^0 \bar{\Delta}^0$, as well as the excited $\Delta$
states and even (excited-)nucleons, is expected to be large. This large interference makes a simple measurement of a two-body decay impossible since the signal yield cannot be extracted by a naive fit with only signal and background components, and need to be considered carefully.

A partial wave analysis (PWA) is necessary to consider this interference correctly, while suitable
formulae to describe these complex processes are still in short of. Till now,
most studies of $\Delta$ is based on $\pi$N or $\gamma N$
scattering~\cite{1810.13086,Gridnev:2004mk,Koch:1980ay,Svarc:2014zja,Bernicha:1995gg},
where the PWA has applied but not suitable for $\psi \to \Delta \bar{\Delta}$ studies since
there are sequent decays in the latter case. Compared with the abundant PWA formulae on
the $\psi$ or heavy vectors decaying into final meson states~\cite{Zou:2002ar, Dulat:2005in}, including radiative decays~\cite{wuning},
the similar results on the baryons are relatively rare. To the best of our knowledge,
there are only two papers that discussed the $\psi$ decays containing final baryon states. One
has studied $\psi \to \omega p \bar{p}$~\cite{Liang:2002tk}, and the other one has studied
$\psi \to N N^*M$~\cite{hep-ph/0210164}, but none of them can satisfy our request
completely. This situation encourages us to write down the PWA formulae by
ourselves.

While there are many different frames and formalism in PWA, we choose to
derive our formulae within the helicity frame with the covariant tensor formalism~\cite{Hara:1964zza, Chung:1993da, Chung:1997jn, Chung:1971ri}. This choice is
based on the following considerations. Firstly, the helicity frame is suitable for processes
of sequent decays. In this frame, all the helicity states are defined in the rest frame
of the mother particles, and the whole amplitude is a product of the amplitudes of sequent processes. It reduces the complexity of the derivation of the formulae compared with the PWA
based on canonical forms, where a L-S coupling method is used such as that in Refs.~\cite{Liang:2002tk, hep-ph/0210164, Filippini:1995yc}. Secondly, the advantage of the tensor formalism, compared to the helicity formalism, is that a spin tensor can couple to any four-momentum or another spin tensor to form the Lorentz invariant amplitudes, and the momentum dependence is separated explicitly from couplings. 
This paper is organized as follows. After this introduction, we discuss all the possible
processes, i.e, the intermediate resonances involved in our partial wave analysis in the
second section. Then we give the Lorentz invariant amplitudes in the third section, composed
of the constructions of each state's wave function, concerns of the parity and charge
conjugation translations, and the explicit covariant tensor formulae.
Finally, we discuss some validations of these formulae and their possible
applications on the present and future electron-positron colliding experiments.

\section{Intermediate resonances and their \texorpdfstring{$J^P$}{}}
\begin{figure}[hbt]
\centering
\includegraphics[width=0.4\textwidth]{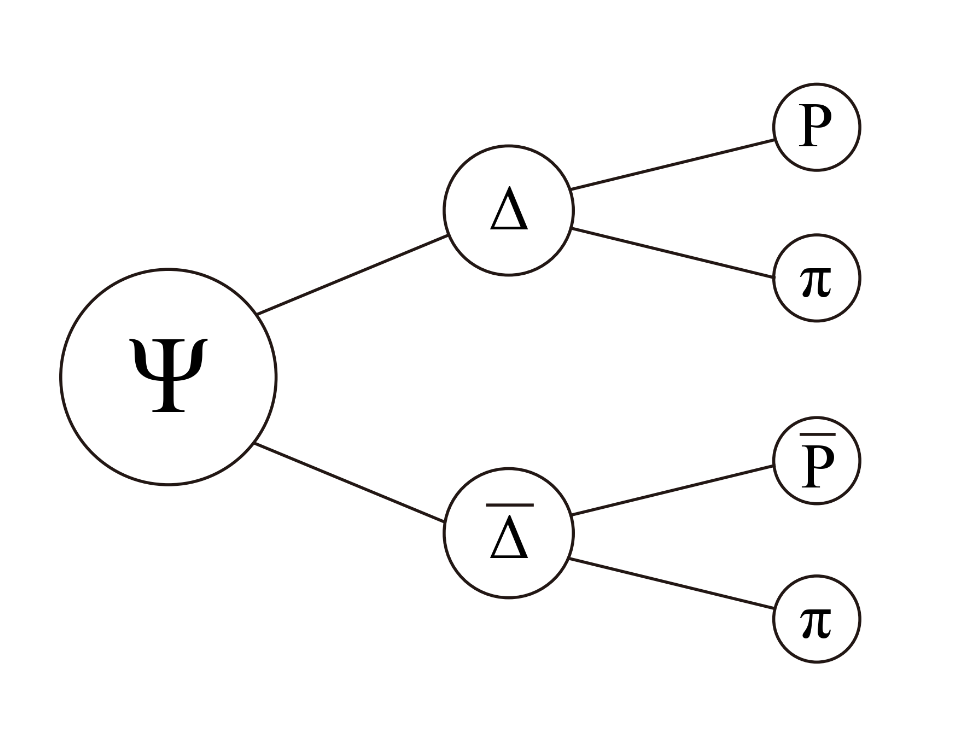}
\caption{Illustration of decay $\psi \to \Delta \bar{\Delta}$, $\Delta \to p \pi$.}
\label{fig:chain1}
\end{figure}

\begin{figure}[hbt]
\centering
\includegraphics[width=0.5\textwidth]{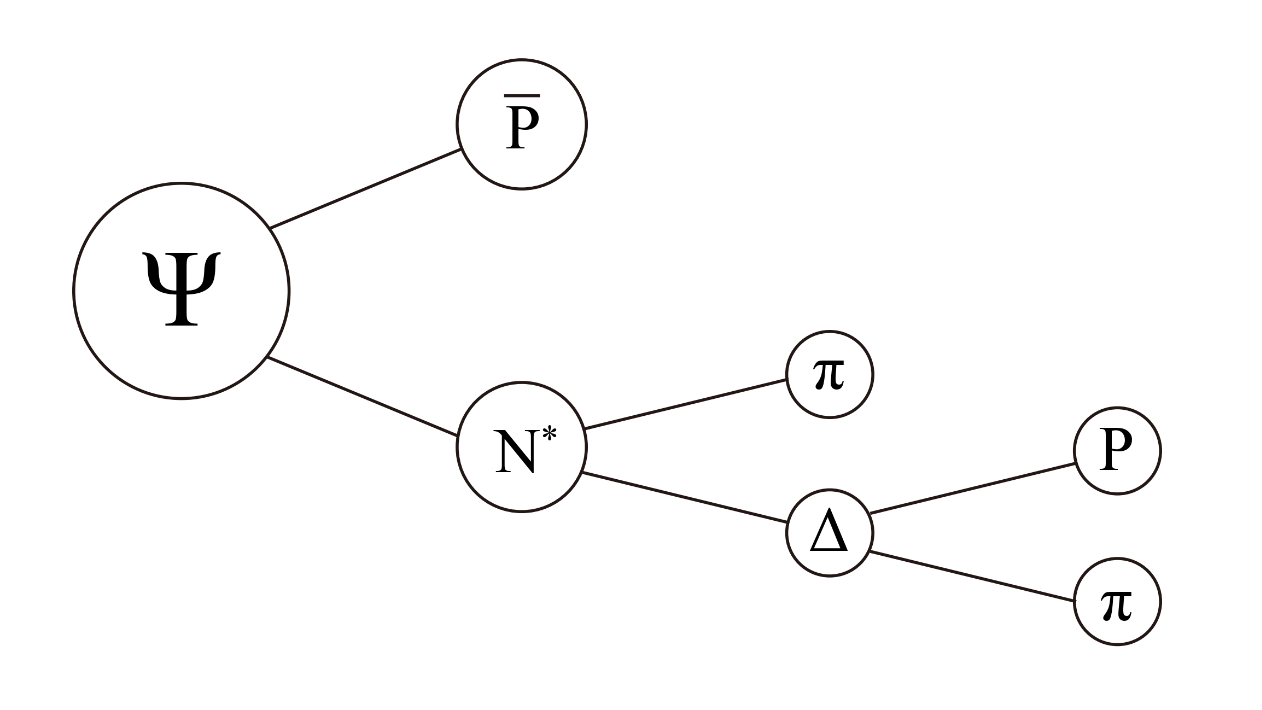}
\caption{Illustration of decay
$\psi \to \bar{p} N^*$, $N^* \to \Delta \pi$, $\Delta \to p \pi$.}
\label{fig:chain2}
\end{figure}

\begin{figure}[hbt]
\centering
\includegraphics[width=0.5\textwidth]{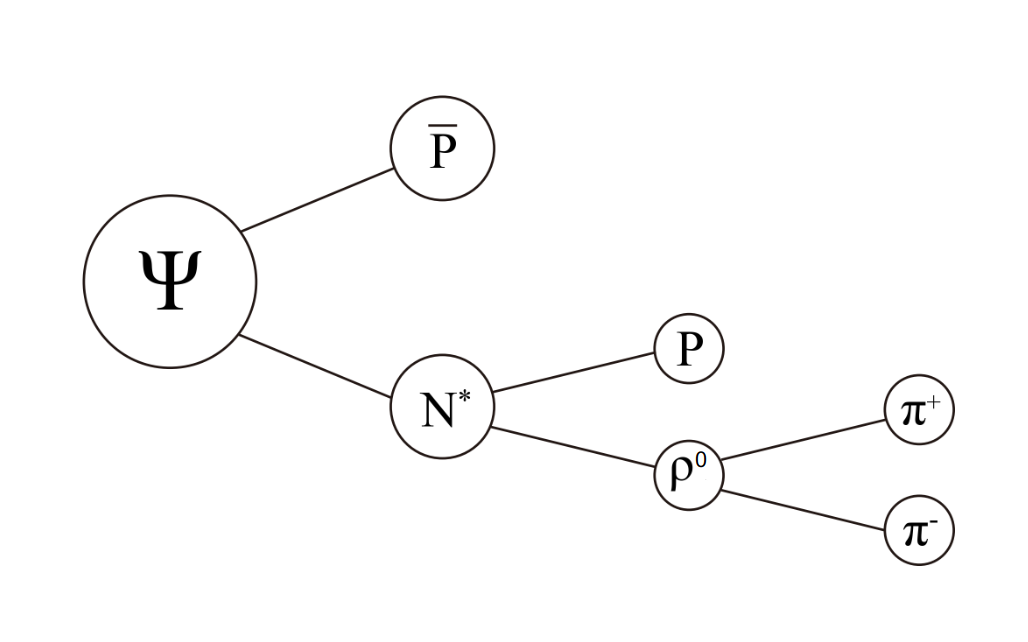}
\caption{Illustration of decay $\psi \to \bar{p} N^*$, $N^* \to p\rho_{0}(\sigma)$, $\rho(\sigma) \to \pi^{+}\pi^{-}$.}
\label{fig:chain3}
\end{figure}

We only consider two kinds of first level decays of $\psi$, that are $\psi \to \Delta
\bar{\Delta}$ and $\psi \to N N^*$. Since our original physics motivation is to measure
$\psi \to \Delta^{++}(1232) \Delta^{--}(1232)$ with $\Delta^{++}(1232)$ almost one hundred
percent decaying to $p$ and $\pi^+$, the suitable final states in experiments are
$p\bar{p} \pi^+ \pi^-$. The final states determine that there are three possible decay
chains, that are $\psi \to \Delta \bar{\Delta}$, $\Delta \to p \pi$;
$\psi \to \bar{p} N^*$, $N^* \to \Delta \pi$, $\Delta \to p \pi$ and
$\psi \to \bar{p} N^*$, $N^* \to p\rho_{0}(\sigma)$, $\rho(\sigma) \to \pi^{+}\pi^{-}$ as shown in Figs.~\ref{fig:chain1}~\ref{fig:chain2}~\ref{fig:chain3}.
The reason why we do not consider the process of $\psi \to \phi X(1870)$, $\phi \to \pi^{+}\pi^{-}, X(1870) \to p\bar{p}$
is that the mass of $X(1870)$ is near the threshold of the $p\bar{p}$ invariant mass spectrum.
Then the relevant events can be removed easily during the date analysis. In the present and
later descriptions we shall omit the discussion on the charge-conjugated modes for compactness when no
ambiguity will be caused. Notice there are many possible combinations even in the first
level decays. For example, $\psi$ can decay into the following $\Delta \bar{\Delta}$ pairs if
the phase space allows: $\Delta(1232)\bar{\Delta}(1620)$, $\Delta(1620)\bar{\Delta}(1620)$,
$\Delta(1232)\bar{\Delta}(1910)$, $\Delta(1620)\bar{\Delta}(1910)$ with $\Delta$ represent
both $\Delta^{++}$ and $\Delta^0$ here. $\Delta^+$ will not be considered since it cannot
decay into both charged p and $\pi$. For the $\psi \to \bar{p} N^*$ mode, the decay
product at the first level can be such as $\bar{p} N(1440)$,  $\bar{p} N(1520)$,  $\bar{p}
N(1675)$, etc.

From the physical consideration, a complex spin and parity combination sets are obtained. At
the first level, decaying from a vector mother particle, i.e. $1^{--}$, the $J^P$ of the decay
products could be $\jp{1}{2}{+}\jp{1}{2}{+}$, $\jp{1}{2}{+}\jp{1}{2}{-}$,
$\jp{1}{2}{+}\jp{3}{2}{+}$, $\jp{1}{2}{+}\jp{3}{2}{-}$, $\jp{3}{2}{+}\jp{3}{2}{+}$, $\jp{3}{2}{+}\jp{3}{2}{-}$,
 $\jp{5}{2}{+}\jp{1}{2}{+}$, $\jp{5}{2}{+}\jp{1}{2}{-}$, 
$\jp{5}{2}{+}\jp{3}{2}{+}$, $\jp{5}{2}{+}\jp{3}{2}{-}$, $\jp{7}{2}{+}\jp{3}{2}{+}$j, and $\jp{7}{2}{+}\jp{3}{2}{-}$ ; at the second level, if the decays are from $\Delta$, the final
states are always $\jp{1}{2}{+} 0^-$ while the $J^P$ of the $\Delta$ can be
$\jp{1}{2}{+}$, $\jp{1}{2}{-}$, $\jp{3}{2}{+}$, $\jp{3}{2}{-}$, $\jp{5}{2}{+}$,
$\jp{5}{2}{-}$, $\jp{7}{2}{+}$, and $\jp{7}{2}{-}$; if a decay chain contains $N^*$ then there is
another decay mode with final states $\Delta \pi$, i.e. $J^P = \jp{3}{2}{+} 0^- $, while
the $J^P$ of the initial $N^*$ can be $\jp{1}{2}{+}$, $\jp{1}{2}{-}$, $\jp{3}{2}{+}$,
$\jp{3}{2}{-}$, $\jp{5}{2}{+}$, $\jp{5}{2}{-}$ $\jp{7}{2}{+}$, and $\jp{7}{2}{-}$;

\section{Preparations}\label{S:prepare}
\subsection{Wave functions}\label{S:wf}
To construct the covariant tensor amplitudes, we need the tensor wave functions describing
relativistic particles with arbitrary spin (helicity) and satisfying the Rarita-Schwinger formalism~\cite{Rarita:1941mf}. We will follow the method proposed by Auvil and Brehm~\cite{Auvil:1966eao} to write down these wave functions explicitly. The wave function of a scalar particle is constant. We start
with a particle of spin-$1$ at rest, whose wave functions may be expressed explicitly as
column vectors:
\begin{equation}
  e(\pm 1) = \mp \frac{1}{\sqrt{2}} \begin{pmatrix} 1 \\ \pm \im \\ 0 \end{pmatrix}, \ \ \
  e(0) =  \begin{pmatrix} 0 \\ 0 \\ 1 \end{pmatrix}.
\end{equation}

Since a particle at rest cannot have the energy component if the spin (helicity) wave function is
presented in the momentum space, the four-vector describing a spin-$1$ particle at rest is
defined as:
\begin{equation}
  e^\mu(0,\lambda) = \{0, \vec{e}(\lambda)\},\ \ \   e_\mu(0,\lambda) = \{0, -\vec{e}(\lambda)\}.
\end{equation}
Then the general helicity state vector with four-momentum $p$ can be obtained by
performing a Lorentz boost in $z$ direction and a proper rotation:
\begin{equation}
  e^\mu(p,0) = \frac{E}{m} \begin{pmatrix} -p/E  \\ \sin\theta \cos\phi  \\ \sin\theta \sin\phi \\ \cos\theta \end{pmatrix},  \ \ \
  e^\mu(p, \pm 1) = \frac{1}{\sqrt{2}} \begin{pmatrix} 0
    \\ \mp \cos\theta \cos\phi + \im \sin\phi  \\ \mp \cos\theta \sin\phi - \im \cos\phi  \\ \pm \sin\theta \end{pmatrix},
\end{equation}
 where $m$, $E$, $p$ and $(\theta,\phi)$ are the invariant mass, energy, momentum, and helicity angles of the particle, respectively, and satisfy $E^2 = p^2 + m^2$ with the natural unit.
 The wave function describing spin-2 particles can be constructed
 out of the wave functions of spin-1 via C-G coefficients as follows:
\begin{equation}
  e^{\mu\nu} (p,2\lambda) = \sum_{\lambda_1 \lambda_2} (1\lambda_1 1 \lambda_2| 2\lambda) e^\mu(p,\lambda_1) e^\nu(p,\lambda_2).
\end{equation}
Then the wave function of spin-$n$ ($n\in \mathrm{I}$) particles can be derived in a
cumulative way as tensor  $e^{\mu_{1}\cdots\mu_{n}}(p, n \lambda)$~\cite{Zhu:1999pu}:
  \begin{equation}
  e^{\mu_{1}\cdots\mu_{n}}(p, n \lambda)=\sum_{\lambda_1,\lambda_2} (n\!\!-\!\!1 \lambda_{1} 1 \lambda_{2}|n\lambda)
  e^{\mu_{1}\cdots\mu_{n-1}}(p,n\!\!-\!\!1 \lambda_1) e^{\mu_n}(p, \lambda_2)
  \end{equation}

Therefore, the wave functions should satisfy the Rarita-Schwinger conditions:
\begin{equation}
p^{\mu_1}e_{\mu_{1}\cdots\mu_{n}}=0
\end{equation}
\begin{equation}
e_{\cdots\mu_i\cdots\mu_j\cdots}=e_{\cdots\mu_j\cdots\mu_i\cdots}
\end{equation}
\begin{equation}
g^{\mu_1\mu_2}e_{\mu_1\cdots\mu_{n}}=0
\end{equation}

We shall turn to the Fermion particles now. Consider a spin-$1/2$ particle at rest and its
basis vectors may be given by the spinors or two-dimensional column vectors:
  \begin{equation}
  \chi(+\frac{1}{2})=  \begin{pmatrix}  1\\  0 \end{pmatrix}, \ \ \
  \chi(-\frac{1}{2})=  \begin{pmatrix}  0\\  1 \end{pmatrix}.
  \end{equation}

We adopt the four-component Dirac spinor to describe the relativistic spin $1/2$
particles. Then the two-component spinors $\chi(\lambda)$ are generalized to the
four-component spinors $u(0,\lambda)$:
  \begin{equation}
  u(0,\lambda)= \begin{pmatrix} \chi(\lambda)\\  0 \end{pmatrix},
  \end{equation}
  where $\lambda=\pm\frac{1}{2}$.
After a boost, the helicity wave functions' explicit expressions can be cast into the form
  \begin{equation}
  u(p,+1/2)=  \frac{1}{\sqrt{2m}}
  \begin{pmatrix}
  \sqrt{E+m} \cos(\theta/2)[\cos(\phi/2)-\im \sin(\phi/2)] \\
  \sqrt{E+m} \sin(\theta/2)[\cos(\phi/2)+\im \sin(\phi/2)] \\
  \sqrt{E-m} \cos(\theta/2)[\cos(\phi/2)-\im \sin(\phi/2)] \\
  \sqrt{E-m} \sin(\theta/2)[\cos(\phi/2)+\im \sin(\phi/2)]
  \end{pmatrix},
  u(p,-1/2)=  \frac{1}{\sqrt{2m}}
  \begin{pmatrix}
  -\sqrt{E+m} \sin(\theta/2)[\cos(\phi/2)-\im \sin(\phi/2)] \\
  \sqrt{E+m} \cos(\theta/2)[\cos(\phi/2)+\im \sin(\phi/2)] \\
  \sqrt{E-m} \sin(\theta/2)[\cos(\phi/2)-\im \sin(\phi/2)] \\
  -\sqrt{E-m} \cos(\theta/2)[\cos(\phi/2)+\im \sin(\phi/2)]
  \end{pmatrix}.
  \end{equation}
The ``adjoint'' spinor is defined as
  \begin{equation}
    \bar{u}(p,\lambda)=u^\dagger (p,\lambda)\gamma^0 \ .
   \end{equation}
The helicity wave functions for the anti-Fermion can be written down similarly as
  \begin{equation}
  v(p,+1/2)=  \frac{1}{\sqrt{2m}}
  \begin{pmatrix}
  \sqrt{E-m} \cos(\theta/2)[\cos(\phi/2)-\im \sin(\phi/2)] \\
  \sqrt{E-m} \sin(\theta/2)[\cos(\phi/2)+\im \sin(\phi/2)] \\
  \sqrt{E+m} \cos(\theta/2)[\cos(\phi/2)-\im \sin(\phi/2)] \\
  \sqrt{E+m} \sin(\theta/2)[\cos(\phi/2)+\im \sin(\phi/2)]
  \end{pmatrix},
  v(p,-1/2)=  \frac{1}{\sqrt{2m}}
  \begin{pmatrix}
  \sqrt{E-m} \sin(\theta/2)[\cos(\phi/2)-\im \sin(\phi/2)] \\
  -\sqrt{E-m} \cos(\theta/2)[\cos(\phi/2)+\im \sin(\phi/2)] \\
  -\sqrt{E+m} \sin(\theta/2)[\cos(\phi/2)-\im \sin(\phi/2)] \\
  \sqrt{E+m} \cos(\theta/2)[\cos(\phi/2)+\im \sin(\phi/2)]
  \end{pmatrix}.
  \end{equation}

Following Auvil and Brehm, wave functions corresponding to particles of spin $j=n+1/2$ can
be constructed by the spin-$1/2$ wave functions and spin-$n$ wave functions with C-G coefficients as the following:
  \begin{equation}
  u^{\mu_{1}\cdots\mu_{n}}(p,j \lambda)=\sum_{\lambda_1 \lambda_2}(n \lambda_1 \frac{1}{2} \lambda_2 | j \lambda)
  e^{\mu_{1}\cdots\mu_{n}}(p, n \lambda_1) u(p, \lambda_2)
  \label{eq:B}
  \end{equation}

The spin-j wave function Eq.~\ref{eq:B} is a four-component spinor with the four-vector indices $\mu_{1}\cdots\mu_{n}$.~Since it describes a state of spin $j$,~it can have only $(2j+1)$ independent components.~The desired supplementary conditions are just the Rarita-Schwinger equations:
\begin{equation}
(\gamma^{\mu}p_{\mu}-m)u_{\mu_1\cdots\mu_n}=0
\end{equation}
\begin{equation}
u_{\cdots\mu_i\cdots\mu_j\cdots}=u_{\cdots\mu_j\cdots\mu_i\cdots}
\end{equation}
\begin{equation}
p^{\mu_1}u_{\mu_1\cdots\mu_n}=0
\end{equation}
\begin{equation}
\gamma^{\mu_1}u_{\mu_1\cdots\mu_n}=0
\end{equation}
\begin{equation}
g^{\mu_1\mu_2}u_{\mu_1\cdots\mu_n}=0
\end{equation}
where $m$ is the mass of the spin-j particle and $p$ is its four-momentum.

\subsection{Effective vertices}\label{S:ev}
We also need effective vertices to construct various partial wave amplitudes. The
principle idea is that the decay mechanism has been considered as effective
interactions, then all the loops in the Feynman diagrams have been absorbed into the
effective vertices~\cite{Scadron:1968zz}. A general form of any amplitude in a single decay chain can be
expressed as
 \begin{equation}
   A=\bar{u}_{1}\Gamma u_{2}eB,
   \label{eq:A}
  \end{equation}
 in which $\bar{u}_1$, $u_2$ and $e$ are wave functions of two baryons and a meson,
 respectively; $B$ is considered as the kernel of the propagator and usually parameterized
 as Breit-Wigner functions; $\Gamma$s are tensors representing effective vertices, who are composed by
 $\mathbb{I}$, $p^{\mu}$, $\gamma^5$, $\gamma^{\mu}$, $\sigma^{\mu\nu}$, $g^{\mu\nu}$,
 $\epsilon^{\mu\nu\alpha\beta}$. The main target of this paper is to find out all the
 independent effective vertices. Before that, it is worthy taking some time to consider
 the constraints on them due to symmetry. Since in these decays, the strong interaction is
 dominant, the conservation is expected under the transformation of parity P and charge
 conjugate C, respectively~\cite{Stapp:1962nxd}. However, $\bar{u}_1$ and $u_2$ do not correspond to precisely a
 particle and its anti-particle, and the charged particles do not have determined C-parity,
 so we only consider the symmetry of parity, and its conservation requires
  \begin{equation}
  \eta^*_1 \eta_2 \eta_A \Gamma_P = 1,
  \end{equation}
where $\Gamma_P$ is the transformation property of different tensors and $\eta_1$,
$\eta_2$, $\eta_A$ are the parities of the two baryons and one meson, respectively.  The properties of different tensors under the P transformation are listed in
Table~\ref{tab:1}. And we also notice the transformation property of wave function
$e^{\mu_1 \cdots \mu_n}$ will be $\prod_{i=1}^{n}(-1)^{\mu_i}$ as $p^\mu$ and
  $\epsilon^{\mu\nu\alpha\beta}$ since each of them only takes Lorentz indexes without the
  Dirac ones.

 \begin{table*}[htbp]
   \renewcommand\arraystretch{1.0}
   \caption{The parity transformation properties of some tensors, the shorthand $(-1)^\mu \equiv 1$ for
     $\mu=0$ while $-1$ for $\mu=1,2,3$ is used as in~\cite{Peskin:1995ev}. }
   \doublerulesep 2pt
   \setlength{\tabcolsep}{2mm}{
     \begin{tabular}{c c c c c c c c}\hline\hline
     &  $\mathbb{I}$ & $\gamma^5$ & $\gamma^\mu$ & $\gamma^\mu \gamma^5$ & $\sigma^{\mu\nu}$
       & $p^\mu$  & $\sigma^{\mu\nu}\gamma^{5}$ \\
     $\Gamma_P$ &  $1$ & $-1$ & $(-1)^\mu$ & $-(-1)^\mu$ & $(-1)^\mu (-1)^\nu$
       & $(-1)^\mu$ & $-(-1)^\mu (-1)^\nu $ \\
       \hline
       \hline
   \end{tabular}}
   \label{tab:1}
 \end{table*}

\section{Decay Amplitudes in tensor formalism} \label{S:probability}
Now we are ready to derive the covariant invariant amplitudes in the tensor formalism for resonance decays. Let us consider a resonance of spin-parity $J^P$ and mass $m$, decaying into two particles 1 and 2:
$$ J \to 1 + 2\ .$$
In the rest frame of the mother resonance, let $p$ is the momentum of the particle 1 with helicity angles $(\theta,\phi)$. Usually the amplitude $A$ describing the decay process into two particles with helicity $\lambda_1$ and $\lambda_2$ may be written as
\begin{equation}\label{eq:hel}
A(\lambda_1, \lambda_2, \theta, \phi) = N_J F^J_{\lambda_1 \lambda_2}D^{J*}_{M,\lambda_1-\lambda_2}(\phi,\theta,0)
\end{equation}
in the helicity formalism, where $N_J$ is the normalization factor, $F^J_{\lambda_1 \lambda_2}$ is the helicity coupling, and $D^{J*}_{M,\lambda_1-\lambda_2}$ is the D function with $M$ is the z-component of the mother's spin. However, with the explicit wave functions and effective vertices, we could construct the independent covariant invariant amplitudes in the tensor formalism.

\subsection{\texorpdfstring{$\psi\rightarrow\Delta\bar{\Delta}$}{}}
 \begin{equation} 
 \begin{split}
   \psi\rightarrow\Delta(\frac{1}{2}^{+})\bar{\Delta}(\frac{1}{2}^{-}): \quad
   &g_1 \bar{u}(p_1,\frac{1}{2}\lambda_1) \gamma^\mu v(p_2,\frac{1}{2}\lambda_2)e_{\mu}(P,\lambda)+   \\
   &g_2 \bar{u}(p_1,\frac{1}{2}\lambda_1) \sigma^{\mu\nu}P_\nu v(p_2,\frac{1}{2}\lambda_2)e_{\mu}(P,\lambda)
 \end{split}
 \end{equation}

  \begin{equation} 
  \begin{split}
    \psi\rightarrow\Delta(\frac{1}{2}^{+})\bar{\Delta}(\frac{1}{2}^{+}): \quad
    &g_1 \bar{u}(p_1, \frac{1}{2}\lambda_1)  \gamma^\mu\gamma^5  v(p_2, \frac{1}{2}\lambda_2)e_{\mu}(P,\lambda)+   \\
    &g_2 \bar{u}(p_1,\frac{1}{2}\lambda_1) \sigma^{\mu\nu}P_\nu\gamma^5 v(p_2, \frac{1}{2}\lambda_2)e_{\mu}(P,\lambda)
  \end{split}
  \end{equation}

  \begin{equation} 
  \begin{split}
    \psi\rightarrow\Delta(\frac{3}{2}^{+})\bar{\Delta}(\frac{1}{2}^{-}): \quad
    &g_1 \bar{u}^\mu(p_1, \frac{3}{2} \lambda_1) v(p_2, \frac{1}{2}\lambda_2)e_{\mu}(P,\lambda)+  \\
    &g_2 \bar{u}_\nu(p_1,\frac{3}{2} \lambda_1)\gamma^{\mu}P^{\nu} v(p_2, \frac{1}{2}\lambda_2)e_{\mu}(P,\lambda)+  \\
    &g_3 \bar{u}_\nu(p_1,\frac{3}{2} \lambda_1)p_{2}^{\mu}P^{\nu} v(p_2, \frac{1}{2}\lambda_2)e_{\mu}(P,\lambda)
  \end{split}
  \end{equation}

  \begin{equation} 
  \begin{split}
    \psi\rightarrow\Delta(\frac{3}{2}^{+})\bar{\Delta}(\frac{1}{2}^{+}): \quad
    &g_1 \bar{u}_\nu(p_1, \frac{3}{2} \lambda_1) \gamma^{5} v(p_2, \frac{1}{2}\lambda_2)e_{\mu}(P,\lambda)+  \\
    &g_2 \bar{u}_\nu(p_1, \frac{3}{2} \lambda_1) \gamma^\mu \gamma^5 P^\nu v(p_2, \frac{1}{2}\lambda_2)e_{\mu}(P,\lambda) +  \\
    &g_3 \bar{u}_\nu(p_1,\frac{3}{2} \lambda_1)p_{2}^{\mu}P^{\nu}\gamma^5 v(p_2, \frac{1}{2}\lambda_2)e_{\mu}(P,\lambda)
  \end{split}
  \end{equation}

  \begin{equation} 
  \begin{split}
    \psi\rightarrow\Delta(\frac{3}{2}^{+})\bar{\Delta}(\frac{3}{2}^{-}): \quad
    &g_1 \bar{u}_\nu(p_1,\frac{3}{2}\lambda_1) \gamma^\mu  v^\nu(p_2,\frac{3}{2}\lambda_2)e_{\mu}(P,\lambda)+  \\
    &g_2 \bar{u}_\nu(p_1,\frac{3}{2}\lambda_1) p_{2}^\mu  v^\nu(p_2,\frac{3}{2}\lambda_2)e_{\mu}(P,\lambda)+  \\
    &g_3 \bar{u}_\nu(p_1,\frac{3}{2}\lambda_1) p_{2}^\nu v^\mu(p_2, \frac{3}{2}\lambda_2)e_{\mu}(P,\lambda)+  \\
    &g_4 \bar{u}_\nu(p_1,\frac{3}{2}\lambda_1) \gamma^\mu P^{\nu} P_{\alpha} v^\alpha(p_2, \frac{3}{2} \lambda_2) e_{\mu} (P,\lambda)+  \\
    &g_5  \bar{u}_\nu(p_1,\frac{3}{2}\lambda_1) p_{1}^\mu P^{\nu} P_{\alpha} v^\alpha(p_2,\frac{3}{2}\lambda_2)e_{\mu}(P,\lambda)
  \end{split}
  \end{equation}

  \begin{equation} 
  \begin{split}
  \psi\rightarrow\Delta(\frac{3}{2}^{+})\bar{\Delta}(\frac{3}{2}^{+}): \quad
    &g_1 \bar{u}_\nu(p_1,\frac{3}{2}\lambda_1) \gamma^\mu\gamma^5  v^\nu(p_2,\frac{3}{2}\lambda_2)e_{\mu}(P,\lambda)+  \\
    &g_2 \bar{u}_\nu(p_1,\frac{3}{2}\lambda_1) p_{2}^\mu\gamma^5  v^\nu(p_2,\frac{3}{2}\lambda_2)e_{\mu}(P,\lambda)+  \\
    &g_3 \bar{u}_\nu(p_1,\frac{3}{2}\lambda_1) p_{2}^\nu \gamma^5 v^\mu(p_2, \frac{3}{2}\lambda_2)e_{\mu}(P,\lambda)+  \\
    &g_4 \bar{u}_\nu(p_1,\frac{3}{2}\lambda_1) \gamma^\mu\gamma^5 P^{\nu} P_{\alpha} v^\alpha(p_2,\frac{3}{2}\lambda_2)e_{\mu}(P,\lambda)+  \\
    &g_5 \bar{u}_\nu(p_1,\frac{3}{2}\lambda_1) p_{1}^\mu P^{\nu} P_{\alpha}\gamma^5 v^\alpha(p_2,\frac{3}{2}\lambda_2)e_{\mu}(P,\lambda)
  \end{split}
  \end{equation}

  \begin{equation} 
  \begin{split}
  \psi\rightarrow\Delta(\frac{5}{2}^{+})\bar{\Delta}(\frac{1}{2}^{-}): \quad
    &g_1 \bar{u}_{\mu\nu}(p_1,\frac{5}{2}\lambda_1) P^\nu v(p_2, \frac{1}{2}\lambda_2)e_{\mu}(P,\lambda)+  \\
    &g_2 \bar{u}_{\nu\alpha}(p_1,\frac{5}{2}\lambda_1) \gamma^{\mu}P^{\nu}P^{\alpha} v(p_2, \frac{1}{2}\lambda_2)e_{\mu}(P,\lambda)+  \\
    &g_3 \bar{u}_{\nu\alpha}(p_1,\frac{5}{2}\lambda_1) p_{1}^{\mu}P^{\nu}P^{\alpha} v(p_2, \frac{1}{2}\lambda_2)e_{\mu}(P,\lambda)
   \end{split}
  \end{equation}

  \begin{equation} 
  \begin{split}
    \psi\rightarrow\Delta(\frac{5}{2}^{+})\bar{\Delta}(\frac{1}{2}^{+}): \quad
    &g_1 \bar{u}_{\mu\nu}(p_1,\frac{5}{2}\lambda_1) P^\nu \gamma^5 v(p_2, \frac{1}{2}\lambda_2)e_{\mu}(P,\lambda)+  \\
    &g_2 \bar{u}_{\nu\alpha}(p_1,\frac{5}{2}\lambda_1) \gamma^{\mu}\gamma^{5}P^{\nu}P^{\alpha} v(p_2, \frac{1}{2}\lambda_2)e_{\mu}(P,\lambda)+  \\
    &g_3 \bar{u}_{\nu\alpha}(p_1,\frac{5}{2}\lambda_1) p_{1}^{\mu}P^{\nu}P^{\alpha}\gamma^5 v(p_2, \frac{1}{2}\lambda_2)e_{\mu}(P,\lambda)
  \end{split}
  \end{equation}

  \begin{equation} 
  \begin{split}
  \psi\rightarrow\Delta(\frac{5}{2}^{+})\bar{\Delta}(\frac{3}{2}^{-}): \quad
    &g_1 \bar{u}^{\mu\nu}(p_1,\frac{5}{2}\lambda_1) v_\nu(p_2,\frac{3}{2}\lambda_2)e_{\mu}(P,\lambda)+  \\
    &g_2 \bar{u}^{\nu\alpha}(p_1,\frac{5}{2}\lambda_1)\gamma^{\mu}P_{\nu} v_\alpha(p_2, \frac{3}{2}\lambda_2)e_{\mu}(P,\lambda)+  \\
    &g_3 \bar{u}^{\nu\alpha}(p_1,\frac{5}{2}\lambda_1)p_{1}^{\mu}P_{\nu} v_\alpha(p_2, \frac{3}{2}\lambda_2)e_{\mu}(P,\lambda)+  \\
    &g_4 \bar{u}^{\nu\alpha}(p_1,\frac{5}{2}\lambda_1)P^{\nu}P_{\alpha} v^\mu(p_2, \frac{3}{2}\lambda_2)e_{\mu}(P,\lambda)+  \\
    &g_5 \bar{u}^{\nu\beta}(p_1,\frac{5}{2}\lambda_1)\gamma^{\mu}P_{\nu}P_{\beta}P_{\alpha} v^\alpha(p_2, \frac{3}{2}\lambda_2)e_{\mu}(P,\lambda)+  \\
    &g_6 \bar{u}^{\nu\beta}(p_1,\frac{5}{2}\lambda_1)p_{1}^{\mu}P_{\nu}P_{\beta}P_{\alpha} v^\alpha(p_2, \frac{3}{2}\lambda_2)e_{\mu}(P,\lambda)  \\
  \end{split}
  \end{equation}

  \begin{equation} 
  \begin{split}
  \psi\rightarrow\Delta(\frac{5}{2}^{+})\bar{\Delta}(\frac{3}{2}^{+}): \quad
    &g_1 \bar{u}^{\mu\nu}(p_1,\frac{5}{2}\lambda_1)\gamma^5 v_\nu(p_2,\frac{3}{2}\lambda_2)e_{\mu}(P,\lambda)+  \\
    &g_2 \bar{u}^{\nu\alpha}(p_1,\frac{5}{2}\lambda_1)\gamma^{\mu}\gamma^{5}P_{\nu} v_\alpha(p_2, \frac{3}{2}\lambda_2)e_{\mu}(P,\lambda)+  \\
    &g_3 \bar{u}^{\nu\alpha}(p_1,\frac{5}{2}\lambda_1)p_{1}^{\mu}P_{\nu}\gamma^5 v_\alpha(p_2, \frac{3}{2}\lambda_2)e_{\mu}(P,\lambda)+  \\
    &g_4 \bar{u}^{\nu\alpha}(p_1,\frac{5}{2}\lambda_1)P^{\nu}P_{\alpha}\gamma^5 v^\mu(p_2, \frac{3}{2}\lambda_2)e_{\mu}(P,\lambda)+  \\
    &g_5 \bar{u}^{\nu\beta}(p_1,\frac{5}{2}\lambda_1)\gamma^{\mu}\gamma^{5} P_{\nu} P_{\beta} P_{\alpha} v^\alpha(p_2, \frac{3}{2}\lambda_2)e_{\mu}(P,\lambda)+  \\
    &g_6 \bar{u}^{\nu\beta}(p_1,\frac{5}{2}\lambda_1)p_{1}^{\mu}P_{\nu}P_{\beta}P_{\alpha}\gamma^5 v^\alpha(p_2, \frac{3}{2}\lambda_2)e_{\mu}(P,\lambda)  
  \end{split}
  \end{equation}

  \begin{equation} 
  \begin{split}
  \psi\rightarrow\Delta(\frac{7}{2}^{+})\bar{\Delta}(\frac{1}{2}^{-}): \quad
   &g_1 \bar{u}^{\mu\nu\alpha}(p_1,\frac{7}{2}\lambda_1)P_{\nu}P_{\alpha} v(p_2, \frac{1}{2}\lambda_2)e_{\mu}(P,\lambda)+  \\
   &g_2 \bar{u}_{\nu\beta\alpha}(p_1,\frac{7}{2}\lambda_1)\gamma^{\mu}P^{\nu}P^{\beta}P^{\alpha} v(p_2, \frac{1}{2}\lambda_2)e_{\mu}(P,\lambda)+  \\
   &g_3 \bar{u}_{\nu\beta\alpha}(p_1,\frac{7}{2}\lambda_1)p_{1}^{\mu}P^{\nu}P^{\beta}P^{\alpha} v(p_2, \frac{1}{2}\lambda_2)e_{\mu}(P,\lambda)
  \end{split}
  \end{equation}

  \begin{equation} 
  \begin{split}
  \psi\rightarrow\Delta(\frac{7}{2}^{+})\bar{\Delta}(\frac{1}{2}^{+}):  \quad
   &g_1 \bar{u}^{\mu\nu\alpha}(p_1,\frac{7}{2}\lambda_1)P_{\nu}P_{\alpha}\gamma^5 v(p_2, \frac{1}{2}\lambda_2)e_{\mu}(P,\lambda)+  \\
   &g_2 \bar{u}_{\nu\beta\alpha}(p_1,\frac{7}{2}\lambda_1)\gamma^{\mu}\gamma^{5}P^{\nu}P^{\beta}P^{\alpha} v(p_2, \frac{1}{2}\lambda_2)e_{\mu}(P,\lambda)+  \\
   &g_3 \bar{u}_{\nu\beta\alpha}(p_1,\frac{7}{2}\lambda_1)p_{1}^{\mu}P^{\nu}P^{\beta}P^{\alpha}\gamma^5 v(p_2, \frac{1}{2}\lambda_2)e_{\mu}(P,\lambda)
  \end{split}
  \end{equation}

  \begin{equation} 
  \begin{split}
  \psi\rightarrow\Delta(\frac{7}{2}^{+})\bar{\Delta}(\frac{3}{2}^{-}): \quad
  &g_1 \bar{u}^{\mu\nu\alpha}(p_1,\frac{7}{2}\lambda_1) P_\nu v_{\alpha}(p_2, \frac{3}{2}\lambda_2)e_{\mu}(P,\lambda)+  \\
  &g_2 \bar{u}_{\nu\beta\alpha}(p_1,\frac{7}{2}\lambda_1) \gamma^{\mu} P^\nu P^{\alpha} v^\beta(p_2, \frac{3}{2}\lambda_2)e_{\mu}(P,\lambda)+  \\
  &g_3 \bar{u}_{\nu\beta\alpha}(p_1,\frac{7}{2}\lambda_1) p_{1}^{\mu} P^\nu P_{\alpha} v^\beta(p_2, \frac{3}{2}\lambda_2)e_{\mu}(P,\lambda)+  \\
  &g_4 \bar{u}_{\mu\beta\alpha}(p_1,\frac{7}{2}\lambda_1) P^{\nu} P^\beta P^{\alpha} v_\nu(p_2, \frac{3}{2}\lambda_2)e_{\mu}(P,\lambda)+  \\
  &g_5 \bar{u}_{\nu\delta\alpha}(p_1,\frac{7}{2}\lambda_1) \gamma^{\mu} P^\nu P^{\delta}P^{\alpha}P^{\beta} v_\beta(p_2, \frac{3}{2}\lambda_2)e_{\mu}(P,\lambda)+  \\
  &g_6 \bar{u}_{\nu\delta\alpha}(p_1,\frac{7}{2}\lambda_1) p_{1}^{\mu} P^\nu P^{\delta} P^{\alpha}P^{\beta} v_\beta(p_2, \frac{3}{2}\lambda_2)e_{\mu}(P,\lambda)  \\
  \end{split}
  \end{equation}

  \begin{equation} 
  \begin{split}
  \psi\rightarrow\Delta(\frac{7}{2}^{+})\bar{\Delta}(\frac{3}{2}^{+}):  \quad
  &g_1 \bar{u}^{\mu\nu\alpha}(p_1,\frac{7}{2}\lambda_1) P_\nu\gamma^5 v_{\alpha}(p_2, \frac{3}{2}\lambda_2)e_{\mu}(P,\lambda)+  \\
  &g_2 \bar{u}_{\nu\beta\alpha}(p_1,\frac{7}{2}\lambda_1) \gamma^{\mu}\gamma^{5} P^\nu P^{\alpha} v^\beta(p_2, \frac{3}{2}\lambda_2)e_{\mu}(P,\lambda)+  \\
  &g_3 \bar{u}_{\nu\beta\alpha}(p_1,\frac{7}{2}\lambda_1) p_{1}^{\mu} P^\nu P_{\alpha}\gamma^5 v^\beta(p_2, \frac{3}{2}\lambda_2)e_{\mu}(P,\lambda)+  \\
  &g_4 \bar{u}_{\mu\beta\alpha}(p_1,\frac{7}{2}\lambda_1) P^{\nu} P^\beta P^{\alpha}\gamma^5 v_\nu(p_2, \frac{3}{2}\lambda_2)e_{\mu}(P,\lambda)+  \\
  &g_5 \bar{u}_{\nu\delta\alpha}(p_1,\frac{7}{2}\lambda_1) \gamma^{\mu}\gamma^{5}P^\nu P^{\delta}P^{\alpha}P^{\beta} v_\beta(p_2, \frac{3}{2}\lambda_2)e_{\mu}(P,\lambda)+  \\
  &g_6 \bar{u}_{\nu\delta\alpha}(p_1,\frac{7}{2}\lambda_1) p_{1}^{\mu} P^\nu P^{\delta} P^{\alpha}P^{\beta}\gamma^5 v_\beta(p_2, \frac{3}{2}\lambda_2)e_{\mu}(P,\lambda)  \\
  \end{split}
  \end{equation}

 \subsection{The formula of \texorpdfstring{$\Delta\to p\pi$}{}}

 \begin{equation} 
 \begin{split}
 \Delta(\frac{1}{2}^{+})\rightarrow p(\frac{1}{2}^{+}) \pi(0^{-}):  \quad
 g \bar{u}(p_{1},\frac{1}{2}\lambda_1)\gamma^{5}u(P,\frac{1}{2}\lambda)
 \end{split}
 \end{equation}

 \begin{equation} 
 \begin{split}
 \Delta(\frac{1}{2}^{-})\rightarrow p(\frac{1}{2}^{+}) \pi(0^{-}):  \quad
 g \bar{u}(p_{1},\frac{1}{2}\lambda_1)u(P,\frac{1}{2}\lambda)
 \end{split}
 \end{equation}

 \begin{equation}
 \begin{split}
 \Delta(\frac{3}{2}^{+})\rightarrow p(\frac{1}{2}^{+}) \pi(0^{-}): \quad
 g \bar{u}(p_{1},\frac{1}{2}\lambda_1)p_{1}^{\mu}\gamma^{5}u_{\mu}(P,\frac{3}{2}\lambda)
 \end{split}
 \end{equation}

  \begin{equation}
 \begin{split}
 \Delta(\frac{3}{2}^{-})\rightarrow p(\frac{1}{2}^{+}) \pi(0^{-}): \quad
  g \bar{u}(p_{1},\frac{1}{2}\lambda_1)p_{1}^{\mu}u_{\mu}(P,\frac{3}{2}\lambda)
 \end{split}
 \end{equation}

 \begin{equation}
 \begin{split}
 \Delta(\frac{5}{2}^{+})\rightarrow p(\frac{1}{2}^{+}) \pi(0^{-}): \quad
 g \bar{u}(p_{1},\frac{1}{2}\lambda_1)p_{1}^{\mu}p_{1}^{\nu}\gamma^{5} u_{\mu\nu}(P,\frac{5}{2}\lambda)
 \end{split}
 \end{equation}

  \begin{equation}
 \begin{split}
 \Delta(\frac{5}{2}^{-})\rightarrow p(\frac{1}{2}^{+}) \pi(0^{-}): \quad
  g \bar{u}(p_{1},\frac{1}{2}\lambda_1)p_{1}^{\mu}p_{1}^{\nu}u_{\mu\nu}(P,\frac{5}{2}\lambda)
 \end{split}
 \end{equation}

 \begin{equation}
 \begin{split}
 \Delta(\frac{7}{2}^{+})\rightarrow p(\frac{1}{2}^{+}) \pi(0^{-}): \quad
 g \bar{u}(p_{1},\frac{1}{2}\lambda_1)p_{1}^{\mu}p_{1}^{\nu}p_{1}^{\alpha}\gamma^{5} u_{\mu\nu\alpha}(P,\frac{7}{2}\lambda)
 \end{split}
 \end{equation}

  \begin{equation}
 \begin{split}
 \Delta(\frac{7}{2}^{-})\rightarrow p(\frac{1}{2}^{+}) \pi(0^{-}): \quad
  g \bar{u}(p_{1},\frac{1}{2}\lambda_1)p_{1}^{\mu}p_{1}^{\nu}p_{1}^{\alpha}u_{\mu\nu\alpha}(P,\frac{7}{2}\lambda)
 \end{split}
 \end{equation}

\subsection{The formula of \texorpdfstring{$\psi\to N^{*}\bar{N}$}{}}
This type of partial wave formulas for $\psi\rightarrow N^{*}\bar{N}$ with the spin of $\bar{N}$ is $\frac{1}{2}^{-}$ has been included in $\psi\rightarrow \Delta\bar{\Delta}$ and will not be repeated.

\subsection{The formula of \texorpdfstring{$N^*\to \Delta\pi$}{}}
\begin{equation}
 \begin{split}
 N^{*}(\frac{1}{2}^{+})\rightarrow \Delta(\frac{3}{2}^{+}) \pi(0^{-}):\quad
 g \bar{u}_{\mu}(p_{1},\frac{3}{2}\lambda_1)P^{\mu}\gamma^{5}u(P,\frac{1}{2}\lambda)
 \end{split}
 \end{equation}

\begin{equation}
\begin{split}
 N^{*}(\frac{1}{2}^{-})\rightarrow \Delta(\frac{3}{2}^{+}) \pi(0^{-}):\quad
 g \bar{u}_{\mu}(p_{1},\frac{3}{2}\lambda_1)P^{\mu}u(P,\frac{1}{2}\lambda)
 \end{split}
 \end{equation}

\begin{equation}
\begin{split}
 N^{*}(\frac{3}{2}^{+})\rightarrow \Delta(\frac{3}{2}^{+}) \pi(0^{-}):\quad
 &g_1 \bar{u}_{\mu}(p_{1},\frac{3}{2}\lambda_1)\gamma^{5}u^{\mu}(P,\frac{3}{2}\lambda)+\\
 &g_2 \bar{u}_{\nu}(p_{1},\frac{3}{2}\lambda_1)p_{1}^{\mu}P^{\nu}\gamma^{5}u_{\mu}(P,\frac{3}{2}\lambda)
 \end{split}
 \end{equation}

\begin{equation}
\begin{split}
 N^{*}(\frac{3}{2}^{-})\rightarrow \Delta(\frac{3}{2}^{+}) \pi(0^{-}):\quad
 &g_1 \bar{u}_{\mu}(p_{1},\frac{3}{2}\lambda_1)u^{\mu}(P,\frac{3}{2}\lambda)+\\
 &g_2 \bar{u}_{\nu}(p_{1},\frac{3}{2}\lambda_1)p_{1}^{\mu}P^{\nu}u_{\mu}(P,\frac{3}{2}\lambda)
 \end{split}
 \end{equation}

\begin{equation}
\begin{split}
 N^{*}(\frac{5}{2}^{+})\rightarrow \Delta(\frac{3}{2}^{+}) \pi(0^{-}):\quad
 &g_1 \bar{u}^{\mu}(p_{1},\frac{3}{2}\lambda_1)p_{1}^{\mu}\gamma^{5}u_{\mu\nu}(P,\frac{5}{2}\lambda)+\\
 &g_2 \bar{u}_{\alpha}(p_{1},\frac{3}{2}\lambda_1)p_{1}^{\mu}p_{1}^{\nu}P^{\alpha}\gamma^{5} u_{\mu\nu}(P,\frac{5}{2}\lambda)
 \end{split}
 \end{equation}

\begin{equation}
\begin{split}
 N^{*}(\frac{5}{2}^{-})\rightarrow \Delta(\frac{3}{2}^{+}) \pi(0^{-}):\quad
 &g_1 \bar{u}^{\nu}(p_{1},\frac{3}{2}\lambda_1)P^{\mu}u_{\mu\nu}(P,\frac{5}{2}\lambda)+\\
 &g_2 \bar{u}_{\alpha}(p_{1},\frac{3}{2}\lambda_1)p_{1}^{\mu}p_{1}^{\nu}P^{\alpha}u_{\mu\nu}(P,\frac{5}{2}\lambda)
 \end{split}
 \end{equation}

\begin{equation}
\begin{split}
 N^{*}(\frac{7}{2}^{+})\rightarrow \Delta(\frac{3}{2}^{+}) \pi(0^{-}):\quad
 &g_1 \bar{u}^{\mu}(p_{1},\frac{3}{2}\lambda_1)p_{1}^{\nu}p_{1}^{\alpha}\gamma^5 u_{\mu\nu\alpha}(P,\frac{7}{2}\lambda) +\\
 &g_2 \bar{u}_{\beta}(p_{1},\frac{3}{2}\lambda_1) p_{1}^{\mu} p_{1}^{\nu} p_{1}^{\alpha} P^{\beta}\gamma^5 u_{\mu\nu\alpha}(P,\frac{7}{2}\lambda)
 \end{split}
 \end{equation}

\begin{equation}
\begin{split}
 N^{*}(\frac{7}{2}^{-})\rightarrow \Delta(\frac{3}{2}^{+}) \pi(0^{-}):\quad
 &g_1 \bar{u}^{\mu}(p_{1},\frac{3}{2}\lambda_1)p_{1}^{\nu}p_{1}^{\alpha}u_{\mu\nu\alpha}(P,\frac{7}{2}\lambda)+\\
 &g_2 \bar{u}_{\beta}(p_{1},\frac{3}{2}\lambda_1)p_{1}^{\mu}p_{1}^{\nu}p_{1}^{\alpha}P^{\beta} u_{\mu\nu\alpha}(P,\frac{7}{2}\lambda)
 \end{split}
 \end{equation}

\subsection{The formula of \texorpdfstring{$N^*\to p\rho^0$}{}}
\begin{equation}
 \begin{split}
 N^{*}(\frac{1}{2}^{+})\rightarrow p(\frac{1}{2}^{+}) \rho^{0}(1^{-}):\quad
 &g_1 \bar{u}(p_{1},\frac{1}{2}\lambda_1) \gamma^{\mu}\gamma^{5}u(P,\frac{1}{2}\lambda) e^{*}_{\mu}(p_{2},\lambda_2)+\\
 &g_2 \bar{u}(p_{1},\frac{1}{2}\lambda_1)P^{\mu}\gamma^{5}u(P,\frac{1}{2}\lambda)e^{*}_{\mu}(p_{2},\lambda_2)
 \end{split}
 \end{equation}

\begin{equation}
 \begin{split}
 N^{*}(\frac{1}{2}^{-})\rightarrow p(\frac{1}{2}^{+}) \rho^{0}(1^{-}):\quad
 &g_1 \bar{u}(p_{1},\frac{1}{2}\lambda_1)\gamma^{\mu}u(P,\frac{1}{2}\lambda)e^{*}_{\mu}(p_{2},\lambda_2)+\\
 &g_2 \bar{u}(p_{1},\frac{1}{2}\lambda_1)P^{\mu}u(P,\frac{1}{2}\lambda)e^{*}_{\mu}(p_{2},\lambda_2)
 \end{split}
 \end{equation}

\begin{equation}
 \begin{split}
 N^{*}(\frac{3}{2}^{+})\rightarrow p(\frac{1}{2}^{+}) \rho^{0}(1^{-}):\quad
 &g_1 \bar{u}(p_{1},\frac{1}{2}\lambda_1)\gamma^{5}u^{\mu}(P,\frac{3}{2}\lambda) e^{*}_{\mu}(p_{2},\lambda_2)+\\
 &g_2 \bar{u}(p_{1},\frac{1}{2}\lambda_1)p_{1}^{\mu}\gamma^{\nu}\gamma^{5}u_{\mu}(P,\frac{3}{2}\lambda) e^{*}_{\nu}(p_{2},\lambda_2)+\\
 &g_3 \bar{u}(p_{1},\frac{1}{2}\lambda_1)p_{1}^{\mu}P^{\nu}\gamma^{5} u_{\mu}(P,\frac{3}{2}\lambda) e^{*}_{\nu}(p_{2},\lambda_2)
 \end{split}
 \end{equation}

\begin{equation}
 \begin{split}
 N^{*}(\frac{3}{2}^{-})\rightarrow p(\frac{1}{2}^{+}) \rho^{0}(1^{-}):\quad
 &g_1 \bar{u}(p_{1},\frac{1}{2}\lambda_1)u^{\mu}(P,\frac{3}{2}\lambda)e^{*}_{\mu}(p_{2},\lambda_2)+\\
 &g_2 \bar{u}(p_{1},\frac{1}{2}\lambda_1)p_{1}^{\mu}\gamma^{\nu}u_{\mu}(P,\frac{3}{2}\lambda) e^{*}_{\nu}(p_{2},\lambda_2)+\\
 &g_3 \bar{u}(p_{1},\frac{1}{2}\lambda_1) p_{1}^{\mu} P^{\nu} u_{\mu}(P,\frac{3}{2}\lambda) e^{*}_{\nu}(p_{2},\lambda_2)
 \end{split}
 \end{equation}

\begin{equation}
 \begin{split}
 N^{*}(\frac{5}{2}^{+})\rightarrow p(\frac{1}{2}^{+}) \rho^{0}(1^{-}):\quad
 &g_1 \bar{u}(p_{1},\frac{1}{2}\lambda_1)p_{1}^{\mu}\gamma^{5}u_{\mu\nu}(P,\frac{5}{2}\lambda) e^{*\nu}(p_{2},\lambda_2)+\\
 &g_2 \bar{u}(p_{1},\frac{1}{2}\lambda_1)p_{1}^{\mu}p_{2}^{\nu}\gamma^{\alpha}\gamma^{5} u_{\mu\nu}(P,\frac{5}{2}\lambda) e^{*}_{\alpha}(p_{2},\lambda_2)+\\
 &g_3 \bar{u}(p_{1},\frac{1}{2}\lambda_1)p_{1}^{\mu}p_{2}^{\nu}P^{\alpha}\gamma^{5}u_{\mu\nu}(P,\frac{5}{2}\lambda) e^{*}_{\alpha}(p_{2},\lambda_2)
 \end{split}
 \end{equation}

\begin{equation}
 \begin{split}
 N^{*}(\frac{5}{2}^{-})\rightarrow p(\frac{1}{2}^{+}) \rho^{0}(1^{-}):\quad
 &g_1 \bar{u}(p_{1},\frac{1}{2}\lambda_1)p_{1}^{\mu}u_{\mu\nu}(P,\frac{5}{2}\lambda)e^{*\nu}(p_{2},\lambda_2)+\\
 &g_2 \bar{u}(p_{1},\frac{1}{2}\lambda_1)p_{1}^{\mu}p_{2}^{\nu}\gamma^{\alpha}u_{\mu\nu}(P,\frac{5}{2}\lambda) e^{*}_{\alpha}(p_{2},\lambda_2)+\\
 &g_3 \bar{u}(p_{1},\frac{1}{2}\lambda_1)p_{1}^{\mu}p_{2}^{\nu}P^{\alpha}u_{\mu\nu}(P,\frac{5}{2}\lambda) e^{*}_{\alpha}(p_{2},\lambda_2)
 \end{split}
 \end{equation}

 \begin{equation}
 \begin{split}
 N^{*}(\frac{7}{2}^{+})\rightarrow p(\frac{1}{2}^{+}) \rho^{0}(1^{-}):\quad
 &g_1 \bar{u}(p_{1},\frac{1}{2}\lambda_1) p_{1}^{\mu}p_{2}^{\nu}u_{\mu\nu\alpha} (P,\frac{7}{2}\lambda) \gamma^{5} e^{*\alpha}(p_{2},\lambda_2)+\\
 &g_2 \bar{u}(p_{1},\frac{1}{2}\lambda_1)p_{1}^{\mu}p_{2}^{\nu} p_{1}^{\beta}\gamma^{\alpha} u_{\mu\nu\beta}(P,\frac{7}{2}\lambda) \gamma^{5}e^{*}_{\alpha}(p_{2},\lambda_2)+\\
 &g_3 \bar{u}(p_{1},\frac{1}{2}\lambda_1) p_{1}^{\mu}p_{2}^{\nu}p_{2}^{\beta} P^{\alpha}u_{\mu\nu\beta}(P,\frac{7}{2}\lambda) \gamma^{5}e^{*}_{\alpha} (p_{2},\lambda_2)
 \end{split}
 \end{equation}

\begin{equation}
 \begin{split}
 N^{*}(\frac{7}{2}^{-})\rightarrow p(\frac{1}{2}^{+}) \rho^{0}(1^{-}):\quad
 &g_1 \bar{u}(p_{1},\frac{1}{2}\lambda_1) p_{1}^{\mu} p_{2}^{\nu}u_{\mu\nu\alpha}(P,\frac{7}{2}\lambda) e^{*\alpha}(p_{2},\lambda_2)+\\
 &g_2 \bar{u}(p_{1},\frac{1}{2}\lambda_1) p_{1}^{\mu} p_{2}^{\nu} p_{1}^{\beta}\gamma^{\alpha} u_{\mu\nu\beta}(P,\frac{7}{2}\lambda) e^{*}_{\alpha}(p_{2},\lambda_2)+\\
 &g_3 \bar{u}(p_{1},\frac{1}{2}\lambda_1) p_{1}^{\mu} p_{2}^{\nu} p_{2}^{\beta} P^{\alpha} u_{\mu\nu\beta}(P,\frac{7}{2}\lambda) e^{*}_{\alpha}(p_{2},\lambda_2)
 \end{split}
 \end{equation}

\subsection{The formula of \texorpdfstring{$N^{*}\to p\sigma$}{}}
\begin{equation}
 \begin{split}
 N^{*}(\frac{1}{2}^{+})\rightarrow p(\frac{1}{2}^{+}) \sigma(0^{+}):\quad
 g \bar{u}(p_{1},\frac{1}{2}\lambda_1)u(P,\frac{1}{2}\lambda)
 \end{split}
 \end{equation}

\begin{equation}
\begin{split}
 N^{*}(\frac{1}{2}^{-})\rightarrow p(\frac{1}{2}^{+}) \sigma(0^{+}):\quad
 g \bar{u}(p_{1},\frac{1}{2}\lambda_1)\gamma^{5}u(P,\frac{1}{2}\lambda)
 \end{split}
 \end{equation}

\begin{equation}
\begin{split}
 N^{*}(\frac{3}{2}^{+})\rightarrow p(\frac{1}{2}^{+}) \sigma(0^{+}):\quad
 g \bar{u}(p_{1},\frac{1}{2}\lambda_1)p_{1\mu}u^{\mu}(P,\frac{3}{2}\lambda)
 \end{split}
 \end{equation}

\begin{equation}
\begin{split}
 N^{*}(\frac{3}{2}^{-})\rightarrow p(\frac{1}{2}^{+}) \sigma(0^{+}):\quad
 g \bar{u}(p_{1},\frac{1}{2}\lambda_1)\gamma^{5}u^{\mu}(P,\frac{3}{2}\lambda)p_{1\mu}
 \end{split}
 \end{equation}

\begin{equation}
\begin{split}
 N^{*}(\frac{5}{2}^{+})\rightarrow p(\frac{1}{2}^{+}) \sigma(0^{+}):\quad
 g \bar{u}(p_{1},\frac{1}{2}\lambda_1)p_{1}^{\mu}p_{1}^{\nu}u_{\mu\nu}(P,\frac{5}{2}\lambda)
 \end{split}
 \end{equation}

\begin{equation}
\begin{split}
 N^{*}(\frac{5}{2}^{-})\rightarrow p(\frac{1}{2}^{+}) \sigma(0^{+}):\quad
 g \bar{u}(p_{1},\frac{1}{2}\lambda_1)p_{1}^{\mu}p_{1}^{\nu}\gamma^{5}u_{\mu\nu}(P,\frac{5}{2}\lambda)
 \end{split}
 \end{equation}

\begin{equation}
\begin{split}
 N^{*}(\frac{7}{2}^{+})\rightarrow p(\frac{1}{2}^{+}) \sigma(0^{+}):\quad
 g \bar{u}(p_{1},\frac{1}{2}\lambda_1)p_{1}^{\mu}p_{1}^{\nu}p_{1}^{\alpha}u_{\mu\nu\alpha}(P,\frac{7}{2}\lambda)
 \end{split}
 \end{equation}

\begin{equation}
\begin{split}
 N^{*}(\frac{7}{2}^{-})\rightarrow p(\frac{1}{2}^{+}) \sigma(0^{+}):\quad
 g \bar{u}(p_{1},\frac{1}{2}\lambda_1)p_{1}^{\mu}p_{1}^{\nu}p_{1}^{\alpha}\gamma^{5} u_{\mu\nu\alpha}(P,\frac{7}{2}\lambda)
 \end{split}
 \end{equation}

\subsection{The formula of \texorpdfstring{$\rho^{0}(\sigma)\to\pi^{+}\pi^{-}$}{}}
For completeness of the given formulae, we give the amplitudes of $\sigma(0^{+})\rightarrow\pi^{+}\pi^{-}$ and $\rho^{0}\rightarrow\pi^{+}\pi^{-}$. The amplitude formula of $\sigma(0^{+})\rightarrow\pi^{+}\pi^{-}$ is 1. The amplitude formula of $\rho^{0}\rightarrow\pi^{+}\pi^{-}$ is
\begin{equation}
\begin{split}
 \rho^{0}(1^{-})\rightarrow \pi^{+}(0^{-}) \pi^{-}(0^{-}):\quad
g p_{1}^{\mu}e_{\mu}(P,\lambda)
 \end{split}
 \end{equation}

\section{Discussion} \label{S:discussion}
\subsection{Comparison to helicity formalism}
Compared with the amplitudes in helicity formalism Eq.~\ref{eq:hel}, it is obvious that the helicity coupling $F^J_{\lambda_1 \lambda_2}$ ought have momentum dependence but not expressed explicitly as what in the tensor formalism. When the measurements deal with narrow resonances such as $\Lambda$ and $\Sigma$, this momentum dependence could be safely ignored. However, when the target resonance is the broad $\Delta$, separating the momentum dependence, i.e. the kinematic effect, from the dynamic part is crucial for the study of the $\Delta$ line-shape that reflects the internal interactions. As a cross check, we have also calculated the number of $F^J_{\lambda_1 \lambda_2}$'s considering relevant symmetry, and it is found to be consistent with the number of $g_i$'s in the tensor formalism.

\subsection{Amplitude for sequent decays}
In the previous section, we have given the sub-amplitudes for each interaction vertex. The whole amplitude of a decay chain then can be obtained straightforward by the product of each sub-amplitudes in the cascading decays. Thus the cross section can be written as 
\begin{equation}
\sigma = \int \mathrm{d}\Omega \left|\cal M \right|^2
\end{equation} with
\begin{equation}
  {\cal M} = \sum_\Lambda \prod_i M_i(\Lambda) = \sum_\Lambda \prod_i BW_i \sum_j g_i^j  A_i^j(\Lambda)\ \, 
  \end{equation}
where $\Omega$ and $\Lambda$ are the helicity angels and the whole configuration of the decay chains and the helicities of each intermediate/final states, respectively. And $BW_i$'s are the propagators describing the dynamic interactions of each decay process and usually in a Breit-Wigner form. Notice in the helicity frame, the direction of a polarization of a state would be different when this state is in different decay chains. So an alignment, usually a rotation, is required to align the directions. This kind of alignment has been handled by experimentalists already~\cite{0806.4098, Aaij:2015tga} and been discussed further by new means in~\cite{Chen:2017gtx, Marangotto:2019ucc, Wang:2020giv}.

\section{Summary} \label{S:summary}
In this paper, we have derived the helicity amplitudes in the tensor formalism needed for the measurement of
$\psi \rightarrow\Delta\bar{\Delta}$, $\Delta \to p \pi$, as well as the formulae for the
sequent decays of the other processes with the same final states. The covariant tensor formalism is adopted instead of the helicity formalism to make
the momentum dependence explicit then more suitable to the measurements with broad resonances such as $\Delta$ and its excited states. The number of independent terms in each decay chain has been checked and confirmed to be same to the number of helicity amplitudes, and of course, same to the number of partial waves in the
decay too. These formulae are prepared for the measurements of $\psi$ decaying into $p \bar{p}
\pi^+ \pi^-$ final states, and also can be extended to the final states such as $p \bar{p}
K^+ K^-$ since their spins and parities are same. Experiments collected large $J/\psi$
and $\psi(2S)$ data samples, such as BESIII, would be benefit from this study.

It should be noticed only P-parity but not C-parity conservation is considered during the
formula derivation since the fermions do not have certain C-parity. But in some special
cases, where the daughter states are particle and anti-particle to each other, the
C-parity is determined to be $(-1)^{L+S}$. It means if the L-S coupling scheme has been
adopted, there is a chance to check the C violation by measuring the amplitude of the
corresponding wave. However, the construction of pure spin wave functions based on high
fractional spin states such as two $3/2$-spin states is beyond our ability. It also should
be noticed that sometimes the momentum notations $P$, $p_1$ and $p_2$ are replaceable in
the formulae since the constructed terms are dependent on each other, so the choice of
them is somewhat just arbitrary. Breit-Wigner functions and projection operators are important in the experimental measurements too. But the projection operators can be obtained by the construction of wave functions mentioned in
this paper straightforwardly. So we ignore the discussion of them. Furthermore, a complete discussion
of the Breit-Wigner functions is beyond this paper's scope, and their choices are
left to experimentalists when some specific forms are needed for specific resonances or
decay channels.

\section*{Acknowledgements}
{We would like to thank Gang Li, ZhenTian Sun, JiaoJiao Song for useful discussion. This work was supported by grants from Natural Science
Foundation of China (No.11735010, U1932108, U1632104, U2032102,
12061131006). 
}

\end{document}